\documentclass[aps,twocolumn,showpacs]{revtex4}
\usepackage{graphics}
\usepackage{graphicx}
\def\bsigma{\mbox{\boldmath $\sigma$}}

\begin{document}
\title{Electrooptics of graphene: field-modulated reflection and birefringence}
\author{M.V. Strikha}
\author{F.T. Vasko}
\email{ftvasko@yahoo.com}
\affiliation{Institute of Semiconductor Physics, NAS of Ukraine, Pr. Nauky 41,
Kyiv, 03028, Ukraine}
\date{\today}

\begin{abstract}
The elecrooptical response of graphene due to heating and drift of carriers is studied
theoretically. Real and imaginary parts of the dynamic conductivity tensor are calculated
for the case of effective momentum relaxation, when anisotropic contributions are small.
We use the quasiequilibrium distribution of electrons and holes, characterized by
the effective temperature of carriers and by concentrations, which are controlled by
gate voltage and in-plane electric field. The geometry of normal propagation of
probe radiation is considered, spectral and field dependences of the reflection
coefficient and the relative absorption are analyzed. The ellipticity degree of the
reflected and transmitted radiation due to small birefringence of graphene sheet
with current have also been determined.
\end{abstract}

\pacs{78.20.Jq, 78.67.Wj, 81.05.ue}

\maketitle

\section{Introduction}
Study of electrooptical response both of bulk semiconductors and of heterostructures
(see Refs. in \cite{1} and \cite{2}, respectively) is a convenient method to characterize
these materials. Such a response is used to modulate both the intensity of radiation
and its polarization. As it was demonstrated more than 30 years ago \cite{3, 4} (see
also Sect. 17 in \cite{5}), the main contribution to the elecrooptical response of
narrow gap semiconductors is caused by the modulation of the interband transitions,
both virtual and real one, under heating and drift of nonequilibrium carriers. The
electrooptical properties of two-dimensional carriers in heterostructures have also
been studied \cite{6}. Since graphene is a gapless semiconductor with linear energy
spectrum \cite{7}, the direct interband transitions in graphene are allowed with the
characteristic interband velocity $v_W=10^{8}$ cm/s, which corresponds the Weyl-Wallace
model \cite{8}, degenerated over spin and valleys. Therefore, optical properties of
graphene should be modulated essentially by temperature and carriers concentration
\cite{9} and these dependences were studied recently. \cite{10} The applied electric
field not only changes carriers temperature and concentration, but also causes
the anisotropy of distribution due to carriers drift \cite{11, 12}. Therefore, the
birefringence effect can be essential for radiation propagating across a graphene sheet
with current. To the best of our knowledge, no measurement of electrooptical response
was performed until recently, and a theoretical study of these phenomena is timely
now.

The results obtained below are based on the tensor of dynamic conductivity, determined
by interband transitions of non-equilibrium carriers. This tensor is determined by
Kubo formula in collisionless approximation \cite{4, 5} with the use of weakly
anisotropic distributions of electrons and holes calculated in \cite{11, 13}. The case
of normal propagation of the incident ($in$), reflected ($r$), and transmitted
($t$) waves of probe radiation (see Fig. 1) is studied, and the reflection and transition
coefficients, controlled by carriers heating conditions, are obtained. It is demonstrated,
that the heating level dependence on applied field, temperature of phonons, and sheet
charge, controlled by gate voltage $V_{g}$, can be verified from electrooptical measurements.
Moreover, graphene is to be considered due to carriers drift as an uniaxial plane, and the
elliptically polarized $r$- and $t$-waves appear under linear polarization of
$in$-radiation, if $\theta \neq 0$ or $\pi/2$, see Fig.1. Due to an effective relaxation
of carriers momenta the distribution anisotropy and the induced birefringence are
small, but a high sensitivity of polarization measurements enables one to determine drift
characteristics of nonequilibrium carriers using a field-induced Kerr effect.

\addvspace{-1 cm}
\begin{figure}[ht]
\begin{center}
\includegraphics{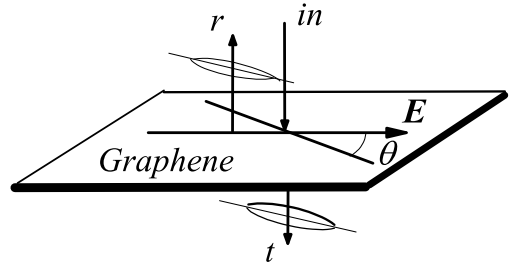}
\end{center}
\addvspace{-1 cm}
\caption{Schematic view on incident ($in$), reflected ($r$), and transmitted ($t$)
radiation for the case of normal propagation through the graphene sheet
with applied electric field $\bf E$. Angle $\theta$ defines the polarization
direction of $in$-wave while $r$-, and $t$-contributions are elliptically
polarized. }
\end{figure}

The consideration below is organised in the following way. In Sec.II we present both
the basic equations for the complex tensor of dynamic conductivity, and the electrodynamics
equations for the uniaxial graphene sheet on a substrate. Numerical results,
describing the electromodulation spectra and Kerr effect, are discussed in Sec.III.
The concluding remarks and the list of assumptions are presented in the last Section.
In Appendix, the dynamic conductivity of the undoped graphene is considered.

\section{Basic equations}
The description of graphene response on the probe in-plane electric field
${\bf E}_{\omega}\exp (-i \omega t)$ is based on the consideration of the
high-frequency dynamic conductivity and on the examination of the electrodynamics
problem for propagation of such a field through graphene sheet. When performing
these calculations, we take into account a modification of interband transitions
due to carriers heating, and an anisotropy of response due to carriers drift.

\subsection{Anisotropic dynamic conductivity}
Contribution of the interband transitions of non-equilibrioum carriers with the
distribution function $f_{l{\bf p}}$, into the response at frequency $\omega$
is described by the dynamic conductivity tensor $\sigma_{\alpha\beta}(\omega )$
given by Kubo formula:
\begin{eqnarray}
\sigma_{\alpha\beta}(\omega )=i\frac{4(ev_W )^2}{\omega L^2}\sum_{ll'{\bf p}}
\left( f_{l{\bf p}}-f_{l'{\bf p}} \right) \\
\times\frac{\left\langle l{\bf p}\left| \hat\sigma_\alpha \right|l'{\bf p}
\right\rangle \left\langle l'{\bf p}\left|\hat\sigma_\beta\right|l{\bf p}
\right\rangle}{\varepsilon_{lp} -\varepsilon_{l'p}+\hbar\omega +i\lambda} . \nonumber
\end{eqnarray}
Here $\left| l{\bf p}\right\rangle$ and $\varepsilon_{lp}$ are the state and
energy of $l$th band ($c$- or $v$-) with the $2D$ momentum $\bf p$, and $\lambda
\rightarrow +0$. We also use the velocity operator $v_{W}\hat{\bsigma}$
and the normalizing area $L^{2}$. The expression (1) is written in a collisionless
approximation $\omega \overline\tau >> 1$ ($\overline\tau$ is the relaxation time),
when intraband transitions are inefficient. In this case the density matrix, averaged
over scattering, should be used in Kubo formula, and $\sigma_{\alpha\beta}(\omega )$
appears to be written through the stationary distribution function $f_{l{\bf p}}$.
\cite{5} Due to effective momentum relaxation the anisotropy of non-equilibrium
electrons and holes distributions is weak and the expansion
\begin{equation}
f_{l{\bf p}}=f_{lp}+\Delta f_{lp}^{(1)}\cos\varphi +\Delta f_{lp}^{(2)}\cos2\varphi
+\ldots
\end{equation}
should be substituted into Eq. (1). Here $\varphi$ angle determines orientation of
$\bf p$ and $\Delta f_{lp}^{(r)}\propto E^{r}$, where ${\bf E}\| OX$ is a dc field
applied. The linear in $E$ contribution can be omitted from $\sigma_{\alpha\beta}
(\omega )$ in the case, when the small spatial dispersion, responsible for the
radiation drag by current (see Ref. 14), is neglected. Thus, with an accuracy of the contributions of $\propto E^{2}$ order, tensor (1) is determined by the transition
matrix elements:
\begin{eqnarray}
\overline{\left|\left\langle 1{\bf p}\left|\hat\sigma_{x,y}\right|
-1{\bf p}\right\rangle\right|^2} = 1/2,  \nonumber \\
\overline{\cos 2\varphi \left| {\left\langle {1{\bf p}\left|\hat\sigma_x
\right| -1{\bf p}}\right\rangle } \right|^2}= -1/4, \\
\overline{\cos 2\varphi \left| {\left\langle {1{\bf p}\left|\hat\sigma_y\right|
-1{\bf p}}\right\rangle }\right|^2}=1/4 , \nonumber
\end{eqnarray}
where overline means the averaging over angle.

Since the non-diagonal components of tensor (1) vanish, the $XX$- and $YY$-components
of dynamic conductivity:
\begin{equation}
\sigma_{xx} (\omega )=\sigma_\omega -\frac{\Delta\sigma_\omega}{2}, ~~~~~
\sigma_{yy} (\omega )=\sigma_\omega +\frac{\Delta\sigma_\omega}{2}
\end{equation}
describe the response of graphene sheet with the field-induced uniaxial anisotropy.
Further, we substitute Eqs. (2) and (3) into (1), and we use the electron-hole
representation, when $ f_{c{\bf p}}\rightarrow f_{e{\bf p}}$, and $f_{v{\bf p}}\rightarrow
1-f_{h,-{\bf p}}$, see \cite{11}. It is convenient to separate the contributions
of the undoped graphene and of the free carriers (electrons and holes) into the
isotropic part of conductivity, $\overline\sigma_\omega$ and $\sigma_\omega^{(c)}$,
so that $\sigma_\omega =\overline\sigma_\omega +\sigma_\omega^{(c)}$. The first
contribution is discussed in the Appendix. Separating the real and imaginary parts
of $\overline\sigma_\omega$, and using the energy conservation law, one obtains
the frequency-independent result ${\rm Re}\overline\sigma =e^2/4\hbar$. The imaginary
contribution into $\overline\sigma_\omega$ is given by the phenomenological
expression (A.2), which is written through the fitting parameters corresponding
to the recent measurements. \cite{15}

The electron-hole contributions to the isotropic part, $\sigma_\omega^{(c)}$,
and the anisotropic addition, $\Delta\sigma_\omega$, are written as follows:
\begin{eqnarray}
{\rm Re}\left|\begin{array}{*{20}c} \sigma_\omega^{(c)}  \\
\Delta\sigma_\omega \end{array}\right| =-\frac{2\pi (ev_W )^2}{\omega L^2}
\sum\limits_{\bf p}\delta (\hbar\omega -2v_Wp) \nonumber \\
\times\left|\begin{array}{*{20}c} f_{ep}+f_{hp}  \\  \Delta f_{ep}^{(2)}+
\Delta f_{hp}^{(2)} \end{array} \right| , ~~~~~
\end{eqnarray}
\begin{eqnarray}
{\rm Im}\left| \begin{array}{*{20}c}\sigma_\omega^{(c)} \\  \Delta\sigma_\omega
\end{array}\right| =-\frac{2(ev_W )^2}{\omega L^2}\sum\limits_{\bf p}
\frac{\cal P}{\hbar\omega -2v_Wp}   \\
\times\frac{4v_W p}{\hbar\omega +2v_Wp}\left|\begin{array}{*{20}c} f_{ep}
+f_{hp} \\ \Delta f_{ep}^{(2)} +\Delta f_{hp}^{(2)} \end{array} \right| .
\nonumber
\end{eqnarray}
The real parts of conductivity given by Eqs. (5) are expressed directly through
isotropic distribution (2) at the characteristic momentum for interband transitions, $p_\omega\equiv\hbar\omega /2v_W$, according to:
\begin{equation}
{\rm Re}\left|\begin{array}{*{20}c}\sigma_\omega^{(c)} \\ \Delta\sigma_\omega  \\
\end{array}\right| =-\frac{e^2}{4\hbar}\left|\begin{array}{*{20}c}
f_{ep_\omega}+f_{hp_\omega} \\  \Delta f_{ep_\omega}^{(2)} +
\Delta f_{hp_\omega}^{(2)}  \\ \end{array} \right| .
\end{equation}
The imaginary parts of $\sigma_\omega^{(c)}$, and $\Delta\sigma_\omega$, given
by Eq. (6) are transformed into:
\begin{eqnarray}
{\rm Im}\left|\begin{array}{*{20}c}\sigma_\omega^{(c)} \\ \Delta\sigma_\omega
\end{array} \right| =-\frac{e^2}{2\pi\hbar p_\omega}\int\limits_0^\infty
\frac{dpp^2}{p_\omega +p}\frac{\cal P}{p_\omega -p} \nonumber  \\
\times\left|\begin{array}{*{20}c} f_{ep} +f_{hp} \\
\Delta f_{ep}^{(2)}+\Delta f_{hp}^{(2)} \end{array} \right|
\end{eqnarray}
and the principal value integrals here should be calculated numerically.

Below, we restrict ourselves to the case of quasielastic distribution of
non-equilibrium electrons and holes ($k=e,h$) with effective temperature
$T_{c}$ and chemical potential $\mu_{k}$:
\begin{equation}
f_{kp}=\{\exp [(v_Wp-\mu_k)/T_c]+1\}^{-1} .
\end{equation}
The dependences of distribution (9) on electric field $E$, temperature
$T$, and gate voltage $V_{g}$ are presented in \cite{11, 13}. For the
anisotropic addition $\Delta f_{k{\bf p}}^{(2)}$ in the case of momentum
relaxation through elastic scattering we use:
\begin{equation}
\Delta f_{kp}^{(2)}=-\frac{(eE)^2p}{2\nu_p^{(2)}}\frac{d}{dp}\left[\frac{1}
{p\nu_p^{(1)}}\left( -\frac{df_{kp}}{dp}\right)\right] .
\end{equation}
For the case of short-range scattering on static defects the relaxation rates
$\nu _p^{(1,2)}$ are proportional to the density of states, so that
$\nu _p^{(1)}=v_d p/\hbar +\nu_0$, and $\nu _p^{(2)}=2\nu _p^{(1)}+\nu_0$. Here
$v_{d}$ is a characteristic velocity, that determines an efficiency of momentum
scattering, \cite{16} and $\nu_0$ is a residual rate, which describes the
scattering process for low-energy carriers.

\subsection{Electrodynamics}
For normal propagation of probe radiation, the Fourier component of the field,
${\bf E}_{\omega z}$, is governed by the wave equation:
\begin{equation}
\frac{d^2 {\bf E}_{\omega z}}{dz^2}+\epsilon_z \left( \frac{\omega}{c}
\right)^2 {\bf E}_{\omega z}+i\frac{4\pi\omega}{c^2}{\bf j}_{\omega z}=0 ,
\end{equation}
where $\epsilon_z$ is dielectric permittivity. In this paper we examine the
case of graphene on the thick SiO$_{2}$ substrate, when $\epsilon_{z<0}=1$,
and $\epsilon_{z>0}= \epsilon$. The induced current density in (11) is localized
around $z=0$ plane, so that ${\bf j}_{\omega z}\approx\hat\sigma_{\omega z}
{\bf E}_{\omega z=0}$, while $\int\limits_{-0}^{+0}dz\hat\sigma_{\omega z}=
\hat\sigma_{\omega}$ is determined through the dynamic conductivity tensor,
being examined above. Outside the graphene sheet the solution of (11) can be
written as:
\begin{equation}
{\bf E}_{\omega z}=\left\{ {\begin{array}{*{20}c}
{{\bf E}^{(in)} e^{ik_\omega  z}  + {\bf E}^{(r)} e^{ - ik_\omega  z} ,} & {z < 0}  \\
{{\bf E}^{(t)} e^{i\overline k_\omega  z} } & {z > 0}  \\
\end{array}} \right. .
\end{equation}
Here the wave vectors $k_\omega =\omega /c$ and $\overline k_\omega =\sqrt{\epsilon}
\omega /c$, and the amplitudes of incident, reflected, and transmitted waves
(${\bf E}^{(in)}$, ${\bf E}^{(r)}$, and ${\bf E}^{(t)}$ respectively) were introduced.
This amplitudes are governed by the boundary condition,
\begin{equation}
\left.\frac{d{\bf E}_{\omega z}}{dz}\right|_{-0}^{+0}+ik_{\omega}\frac{4\pi}{c}
\hat\sigma_\omega{\bf E}_{\omega z=0}=0,
\end{equation}
which is obtained after integration of Eq.(11) over $z$ through the graphene
layer $(-0<z<+0)$. The second boundary condition is the requirement of continuity:
${\bf E}_{\omega z=-0}={\bf E}_{\omega z=+0}$.

Taking into account the diagonality of $\hat\sigma_{\omega}$ tensor,
we get the solutions from the boundary conditions as follows:
\begin{equation}
E_\alpha^{(t)}=\frac{2}{1+A_\alpha (\omega )}E_\alpha^{(in)},
~~~~ E_\alpha^{(r)}=\frac{1-A_\alpha (\omega )}{1+A_\alpha (\omega )}E_\alpha^{(in)} ,
\end{equation}
where factor $A_\alpha (\omega )=\sqrt\epsilon +4\pi\sigma_{\alpha\alpha}(\omega )/c$
was introduced. After substitution of Eqs. (12) and (14) into the general expression
for Poynting vector ${\bf S}=c^2{\rm Re}\left[{\bf E}\times {\rm rot}{\bf E}^*\right]
/8\pi$, we obtain the incident, reflected, and transmitted fluxes $S_{in}$, $S_r$,
and $S_t$ respectively, which are parallel to $OZ$. After multiplication of Eq. (13)
by ${\bf E}_t^*$, we get the relation between these fluxes as follows:
\begin{equation}
S_{in}=S_r+S_t+\frac{1}{2}{\rm Re}\left({\bf E}_t^*\cdot\hat\sigma_\omega\cdot
{\bf E}_t\right) ,
\end{equation}
where the last term describes absorption.

The polarization characteristics of $r$-, and $t$-waves are determined by solutions (14).
It is convenient to present them in complex form $E_\alpha ={\cal E}_\alpha \exp
(i\psi_\alpha )$, where ${\cal E}_\alpha$, and $\psi_\alpha $ give the amplitude and
phase of $\alpha$-component of the field respectively. At $\theta = 0$, or at $\theta
= \pi /2$, when the response is described by $\sigma_{xx}$, or by $\sigma_{yy}$, the
linearly modulated $r$-, and $t$-waves occur. For other $\theta$, the reflected
and transmitted waves are elliptically polarized. The ellipticity degree
$\varepsilon (\omega )$ is determined by the phases difference between $X$- and
$Y$-components of the field, $\Delta\psi =\psi_x-\psi_y$, see the general expressions
in Ref. 17. Under weak anisotropy, with the accuracy up to first order in
$\Delta\sigma_{\omega}$, we get $\varepsilon (\omega )=\Delta\psi /2$.

\begin{figure}
\begin{center}
\includegraphics{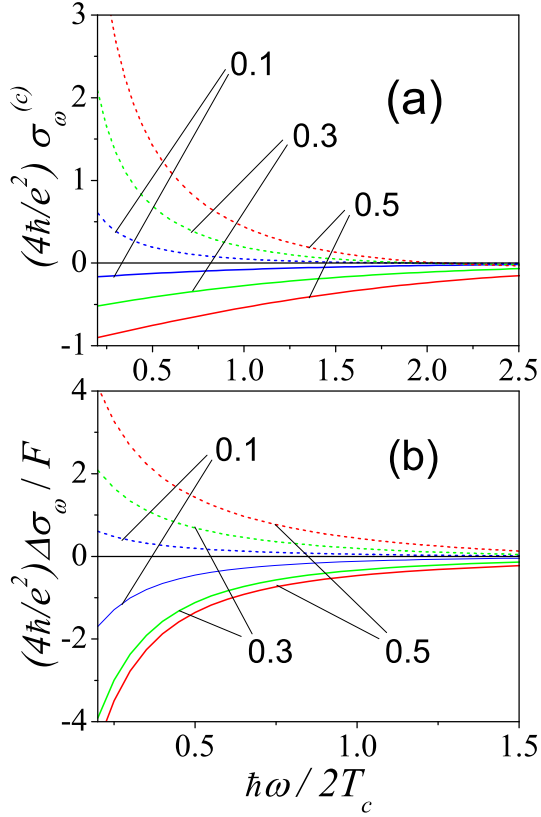}
\end{center}
\addvspace{-1 cm}\caption{Spectral dependences
of $\sigma_\omega^{(c)}$ (a) and $\Delta\sigma_\omega$ (b) for intrinsic graphene
with $f_{p=0}=$0.5, 0.3, and 0.1. Solid and dashed curves correspond real and
imaginary parts of conductivity, respectively.}
\label{fig.2}
\end{figure}

\section{Results}
Now we examine the spectral and polarization characteristics of the
electrooptical response. We study the reflection, transmission, and relative
absorption coefficients, determined as $R_{\omega\theta}=S_r/S_{in}$,
$T_{\omega\theta}=S_t/S_{in}$, and $\xi_\omega ={\rm Re}\left({\bf E}_t^*
\cdot\hat\sigma_\omega\cdot{\bf E}_t\right)/2S_{in}$, respectively, \cite{18}
as well as the ellipticity degree, $\varepsilon (\omega )$, for the case of
weak anisotropy. The final expressions for the coefficients under consideration
are obtained with the use of complex conductivities $\sigma_{\omega}^{(c)}$, and
$\Delta\sigma_{\omega}$, given by Eqs.(7), and (8), and they depend both on
$\hbar\omega /T_{c}$, and on carriers concentration. In Fig.2. we plot these
dependences and one can see that the response modify essentially with
temperature and concentration. The smallness of anisotropic additions is
determined by dimensionless factor
\begin{equation}
F=\left(\frac{eE\hbar v_W^2}{2T_c^2v_d}\right)^2 ,
\end{equation}
which arises from $\propto E^2$ contribution to the distribution function (10).
Note also, that ${\rm Im}\Delta\sigma_{\omega}$ depends weakly on the cutting
parameter $(\hbar\nu_0v_W)/(T_cv_d)$, taken below as 0.1.

\subsection{Reflection and absorption}
For the examination of $R_{\omega\theta}$, and $\xi_{\omega\theta}$ it is convenient
to separate the isotropic and $\theta$-dependent contributions, so that
\begin{equation}
R_{\omega\theta}=R_\omega +\Delta R_\omega\cos 2\theta , ~~~~
\xi_{\omega\theta}=\xi_\omega +\Delta\xi_\omega\cos 2\theta ,
\end{equation}
where the small (of the order of $\Delta\sigma_{\omega}/\sigma_{\omega}$)
anisotropic additions, proportional to $\cos 2\theta$, have been separated,
see Fig. 1. The coefficients in Eq. (17) are written below through
$\sigma_\omega$, $\Delta\sigma_\omega$ and the factor $A_\omega =\sqrt\epsilon
+4\pi\sigma_\omega /c$. For the isotropic parts of reflection, and relative absorption
coefficiens we get \cite{9}:
\begin{equation}
R_\omega\simeq\left|\frac{1-A_\omega}{1+A_\omega}\right|^2 , ~~~~
\xi_\omega\simeq\frac{16\pi}{\sqrt\epsilon c}\frac{{\rm Re}\sigma_\omega}
{|1+A_\omega |^2} ,
\end{equation}
so that these characteristics depend on $T$, $E$, and $V_g$.
\begin{figure}
\begin{center}
\includegraphics{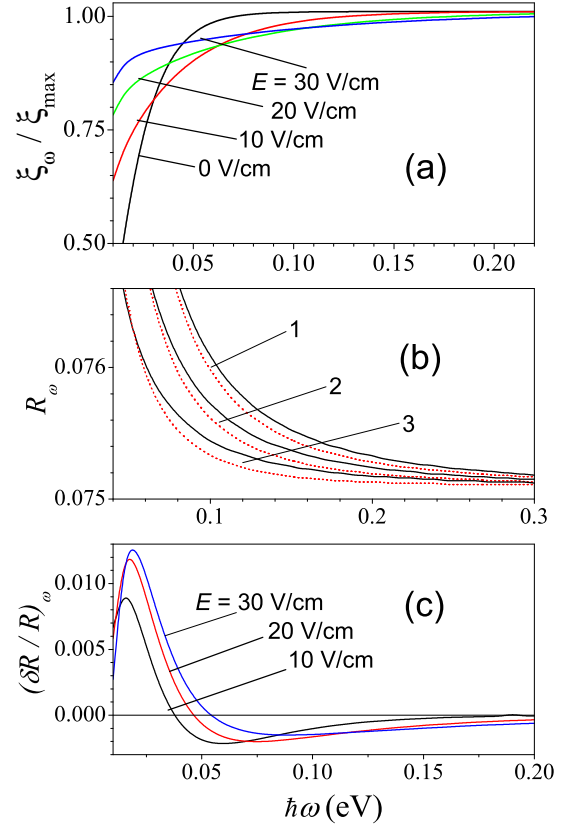}
\end{center}
\addvspace{-1 cm}\caption{Spectral dependencies of relative absorption (a),
reflection (b) and differential reflectivity (c) for intrinsic graphene at
77 K and at different electric fields, $E$ (marked). Solid and dashed
curves in (b) are plotted at $E=$0 and 30 V/cm, respectively, for
$\varepsilon_m=$60 meV (1), 80 meV (2), and 100 meV (3).}
\label{fig.3}
\end{figure}

Spectral dependences of the relative absorption, $\xi_{\omega}/\xi_{max}$,
the reflection, $R_{\omega}$, and the differential reflectivity
$(\delta R/R)_{\omega}\equiv (R_{\omega}-R_{\omega}^{(eq)})/R_{\omega}^{(eq)}$,
are plotted in Fig.3 for intrinsic graphene at $T=$ 77 K and different electric
fields (the data for $T_c$ and carriers concentration were used from Ref. 11).
Here $\xi_{max}$ is the maximum value of relative absorption for high frequences,
when the free carriers contribution is unessential. One can see, that due to
the increase of average energy of carriers with the increase of  $E$ the
absorption increases at high frequencies and decreases for the low ones. The
relative change of $\xi_{\omega}$ is reasonably large, and for
$\hbar\omega\sim T_c$ it can be measured directly. At the same time, the
reflection coefficient depends on field in more weak way, see Figs. 3b, c,
and $(\delta R/R)_{\omega}$ can be 10$^{-2}$ in THz spectral region; in
near-IR spectral region it decreases down to value $\leq 10^{-3}$.
Note, that for $\hbar\omega\leq$0.1 eV $R_{\omega}$ increases essentially
(at high frequencies $R_{\omega}\sim$0.075) due to the contribution of the first
summand of Eq. (A2). Fig. 3b presents the dependence of $R_{\omega}$ on
phenomenological parameter $\varepsilon_m$; $\xi_{\omega}$ depends weakly on
this parameter.

\begin{figure}
\begin{center}
\includegraphics{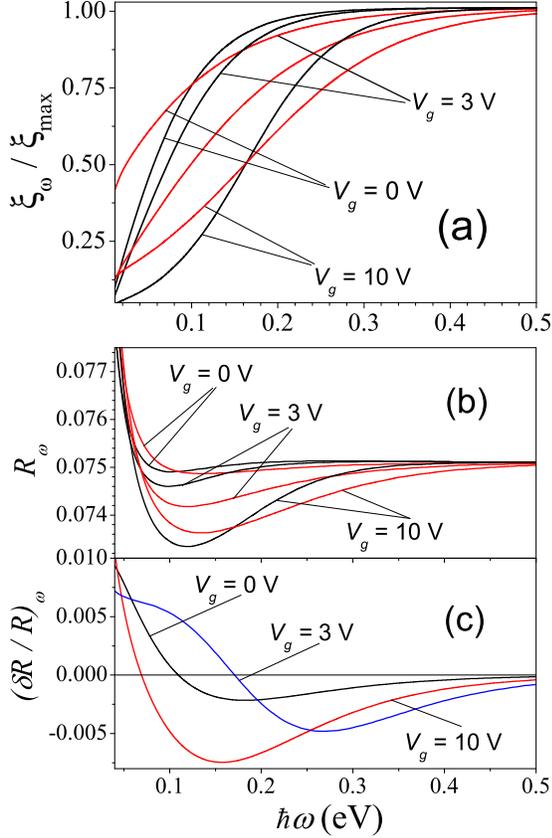}
\end{center}
\addvspace{-1 cm}
\caption{The same as in Fig. 3 for doped graphene at room temperature and
different $V_g$ (marked). Solid and dashed curves correspond $E=$0
and 30 V/cm, respectively.}
\label{fig.4}
\end{figure}

The dependences of $\xi_{\omega}$, $R_{\omega}$, and $(\delta R/R)_{\omega}$ on
doping level are presented in Fig.4. The data for the room temperature are
presented for $V_g=$3 V and 10 V, which correspond the difference between
electron and hole concentrations $1.65\times 10^{11}$ cm$^{-2}$ and
$5.5\times 10^{11}$ cm$^{-2}$, respectively. Similarly to field dependences
at $T=$77 K (see Fig.3), with the increase of $V_g$ (the doping level $\propto V_g$)
the response moves towards the high energy region. The dependences on the level
of heating (the applied field $E$), and on carriers concentration (the gate
voltage $V_g$) correspond the measurements of spectra for different
temperatures and $V_g$, see \cite{10}. For the range of parameters under
examination the field modulation of $\xi_\omega$ is of 20$\div$50 \% order up
to mid-IR spectral region. These modifications should be observed rather easily.
The carriers contribution into reflection increases as well: at $\hbar\omega >$0.1 eV
the decrease of $R_{\omega}$ occurs, which almost does not depend on $\varepsilon_m$;
in this case the shape of the differential reflectivity is similar to
low temperature case, with the shift into the high energy region.

\begin{figure}
\begin{center}
\includegraphics{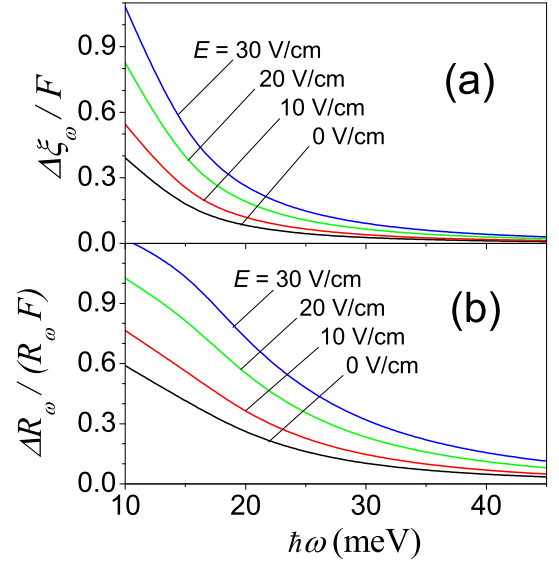}
\end{center}
\addvspace{-1 cm}
\caption{Spectral dependences of anisotropic contributions to relative absorption,
$\Delta\xi_\omega /F$ (a), and to reflectivity, $\Delta R_\omega /(R_\omega F)$,
(b) for intrinsic graphene at $T=$300 K and at different electric fields, $E$.}
\label{fig.5}
\end{figure}

Later we shall examine the anisotropic contributions in Eq.(17), which are
proportional to $\Delta \sigma_{\omega}$. Such a contribution into reflection
coefficient is given by:
\begin{equation}
\Delta R_\omega =\frac{R_\omega}{c}\left(\frac{4\pi\Delta\sigma_\omega}{1-
A_\omega^2}+c.c. \right)
\end{equation}
and the addition to relative absorption takes form:
\begin{eqnarray}
\Delta\xi_\omega =\frac{16\pi}{\sqrt\epsilon c}\frac{{\rm Re}\sigma_\omega}
{|1+A_\omega |^2}\left(\frac{2\pi\Delta\sigma_\omega /c}{1+A_\omega} + c.c.\right)
\nonumber \\
-\frac{16\pi}{\sqrt\epsilon c}\frac{{\rm Re}\Delta\sigma_\omega}{|1+A_\omega |^2} .
\end{eqnarray}
Spectral dependences for anisotropic contributions to the relative absorption and
reflectivity, $\Delta\xi_{\omega}/F$ and $\Delta R_{\omega}/(R_\omega F)$, are shown
in Figs. 5 (a) and (b). One can see, that in the range of fields under examination
the parameter $F$ given by Eq. (16) does not exceed 0.05, so that $\Delta\xi_{\omega}$
and $\Delta R_{\omega}/R_\omega$ are of $10^{-4}$ order for the mid-IR spectral
region ($\sim 0.1\div$0.2 eV) and the response increases up to $\sim 10^{-2}$
in THz spectral region. The anisotropy of such order of value can be analyzed with
the use of the modulation methods.

\subsection{Kerr effect}
Besides the cases of parallel or transverse orientation of the probe radiation
polarization with respect to the drift direction (i. e. at $\theta\neq 0,~ \pi/2$),
the reflected and transmitted fields are elliptically polarized. The maximal
Kerr effect occurs if the $i$-wave is polarized along $\theta =\pi /4$, and
below we consider this case only. In the approximation of the weakly anisotropic
distribution (2) the ellipse orientation does not differ essentially from
$\theta\simeq\pi /4$, and the ellipticity degree can be written as \cite{17}:
\begin{equation}
\varepsilon (\omega )=\sin\beta{\rm Re}\Phi (\omega )-\cos\beta{\rm Im}
\Phi (\omega ) .
\end{equation}
Here the $\beta$ angle, and the complex function $\Phi (\omega )$ can be expressed
through the difference of the phases of $r$-, and $t$-waves (see the end of Sec.II).
The smallness of the ellipticity is determined by the relation $\Phi (\omega )
\propto F$, while the $\beta$ angle is not small.

For the reflected wave the $\Phi (\omega )$ function is given by the expression
\begin{equation}
\Phi_r (\omega )=\frac{4\pi\Delta\sigma_\omega}{c\left( 1+A_\omega \right)^2}
\left|\frac{1+A_\omega}{1-A_\omega} \right| ,
\end{equation}
while the $\beta$ angle is introduced through the relation:
\begin{equation}
\tan \beta_r =-\frac{2{\rm Im}A_\omega}{1-|A_\omega |^2} .
\end{equation}
Similarly, for the transmitted wave, (20) is expressed through the function:
\begin{equation}
\Phi_t (\omega ) = \frac{2\pi\Delta\sigma_\omega}{c(1+A_\omega )^2}
|1 + A_\omega |
\end{equation}
and the $\beta$ angle is given by the expression:
\begin{equation}
\tan\beta_t=-\frac{{\rm Im}A_\omega}{1+{\rm Re}A_\omega} .
\end{equation}
Substitution of these expressions into Eq. (21) gives the ellipticity degrees
for $r$- and $t$-waves, $\varepsilon_r (\omega )$ and $\varepsilon_t (\omega )$.

\begin{figure}
\begin{center}
\includegraphics{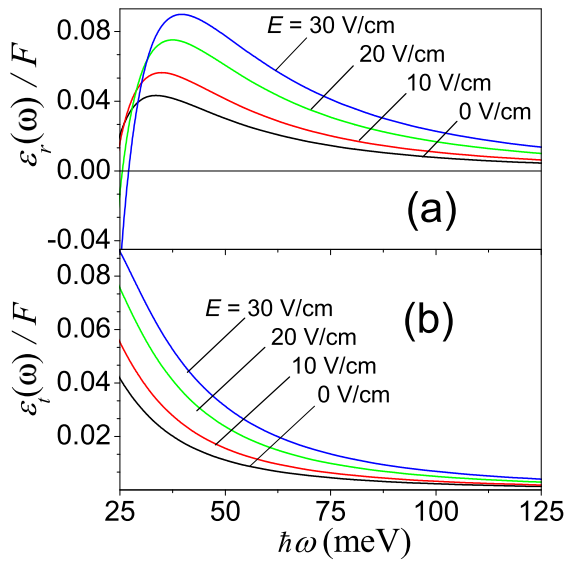}
\end{center}\addvspace{-1 cm}
\caption{Spectral dependences of ellipticity degrees of $r$- and $t$-waves
[panels (a) and (b), respectively] for intrinsic graphene at $T=$300 K and
different electric fields, $E$. }
\label{fig.6}
\end{figure}

Spectral and field dependences of $\varepsilon_r (\omega )$ and $\varepsilon_t
(\omega )$ are shown in Figs. 6a and 6b for intrinsic graphene at $T=$300 K.
In mid-IR spectral region $\varepsilon_{r,t} (\omega )$  decreases with
$\omega$ and for $\hbar\omega\sim 0.1\div$0.2 eV the value of ellipticity
degree does not exceed $\sim 10^{-4}$ at $F\sim$0.05. In THz spectral
region $\varepsilon_{r,t}(\omega )$ increases up to $\sim 5\cdot 10^{-3}$,
wherein the direction of rotation for the reflected wave changes at
$\hbar\omega\sim$25 meV. Such value of ellipticity degree can be detected by
modulation methods only. However, in stronger fields, when the distribution
function is strongly anisotropic, \cite{12} the ellipticity degree can
increase essentially.

\section{Conclusions}

Summarizing the consideration performed, we have examined the graphene
electooptical response due to the interband electron transitions under the
carriers heating and drift. It was found, that an essential modulation
of the reflection, and the relative absorption take place starting from the
field strength $\sim$30 V/c at liquid nitrogen and room temperatures (with the
increase of field, the modulation should increase essentially). Due to
current-induced birefringence of graphene sheet the weak ellipticity of the
reflected and transmitted radiation arise.

Next, we list and discuss the assumptions used in our calculations. First, the
dynamic conductivity tensor (1) is written in collisionless approximation.
For the case of short-range scattering, when $\omega >> v_{d}p_{\omega}/\hbar$,
one arrives to the condition $v_{d}/2v_W\ll 1$ and the collisionless approximation
is not valid for a strongly disordered material. Also, the interband response of
a pure graphene is described with the use of the phenomenological expression
(A.2) and a low-frequency restriction for this approximation is not clear.

Second, the quasiequilibrium distribution of carriers (9) was used for the
numerical estimation of electrooptical response. This means the assumption
of an effective intercarrier scattering. The complete description of the
carriers heating under such conditions had not been performed yet \cite{11,19}.
However, the approximation (9) gives a good estimation for the response
magnitude, and the peculiarities of spectral dependences enable us to
determine the contributions of the different relaxation mechanisms. Similarly,
the use of short-range scattering model in the drift-induced contribution (10)
gives the estimation for optical anisotropy magnitude, and the spectral
dependences peculiarities contain information about the momentum relaxation
mechanism (despite the short-range scattering can be treated as a dominant one
within the phenomenological description of momentum relaxation, \cite{16}
the microscopic mechanism have not been verified until now \cite{20}).

Third, we have examined the heating of carriers with low energies only,
(the results for $E\leq$30 V/cm have been presented), when the essential
electrooptical response occurs in THz spectral region only. With the increase
of field (up to tens kV/cm, see \cite{12}) the electrooptic effect increases and
shifts into near-IR spectral region. The theoretical approach developed here can
be applied for this region as well, however the calculation of the distribution
of hot carriers for this case have not been performed yet.

Forth, the case of graphene on a thick substrate have been examined. The
consideration of the interference effects for graphene, placed on substrate
of limited thickness, needs more complicated calculations (however, an accuracy
of measurements can increase for near-IR spectral region [21]), and is beyond the
frame of this paper. And the last, we have limited ourselves to the examination
of the geometry of normal propagation of radiation only. The study of the response
dependence on the angle of radiation falling gives additional experimental data,
however it is more complicated and needs special treatment.

Finally, the results obtained demonstrate, that the electrooptical response due
to heating and drift of carriers is large enough, and it can be measured.
Because of strong dependence of the response on the applied field, temperature, and
gate voltage, these measurements can give an information on relaxation and
recombination mechanisms. In addition, the electrooptical response of
graphene can be applied for modulation of intensity and polarization
of radiation in THz and mid-IR spectral regions.

\appendix*
\section{Response of undoped graphene}
The dynamic conductivity for the case of undoped graphene is described by
Eqs. (1) and (3) after replacement of $f_{v{\bf p}}$ by 1 and of $f_{c{\bf p}}$
by 0. As a result we get the expression:
\begin{equation}
\overline\sigma_\omega =\frac{2(ev_W )^2}{\omega L^2}\sum\limits_{\bf p}
\left[\pi\delta (\hbar\omega -2v_Wp)+\frac{i\cal P}{\hbar\omega -2v_Wp}
\right] ,
\end{equation}
where the real and imaginary parts of conductivity have been separated. The
direct integration with the use of the energy conservation law gives the
frequency-independent real part of (A1): ${\rm Re}\overline\sigma =e^2/4\hbar$.
The imaginary contribution into $\overline \sigma_\omega$ appears to be
divergent at $p\to \infty$; moreover ${\rm Im}\overline\sigma_\omega\propto
v_Wp_m/\hbar\omega$, where $p_m$ is a cut-off momentum. \cite{22} On the contrary
to the case of bulk material, \cite{9} this cut-off appears to be too
rough for the description of the response in graphene. It is convenient to
approximate $Im \overline \sigma_\omega$, by separating the terms $\propto
\omega^{-1}$, and $\propto\omega$, which correspond the contributions of
the virtual interband transitions, and of ions, correspondingly. As a result,
we get:
\begin{equation}
{\rm Im}\overline\sigma_\omega\approx\frac{e^2}{\hbar}\left(\frac{\varepsilon_m}
{\hbar\omega}-\frac{\hbar\omega}{\varepsilon_i}\right) ,
\end{equation}
where the characteristic energies, $\varepsilon_m$, and $\varepsilon_i$, have been
introduced. The comparison of the response, described by $\overline \sigma_\omega$,
with the recent measurements of the graphene optical spectra, yields:
$\varepsilon_m \sim$0.08 eV, and $\varepsilon_i \sim$ 6.75 eV. \cite{15}


\begin{thebibliography}{52}
\bibitem{18}
According to Eq. (15), the reflection, transmission, and relative absorption
coefficients are connected by the relation:
$1=R_{\omega\theta}+T_{\omega\theta}+\xi_{\omega\theta}$.
\end{thebibliography}
\end{document}